\documentclass[aps,pra,twocolumn,amsmath,amssymb,nofootinbib,showpacs,superscriptaddress]{revtex4-1}
\usepackage[english]{babel}
\usepackage{latexsym}
\usepackage{graphics}
\usepackage{braket}
\usepackage{graphicx}
\usepackage{epsfig}
\usepackage{color}
\usepackage{bm}
\usepackage{amsmath}
\usepackage{amssymb}
\usepackage{amsthm}
\usepackage{dcolumn}
\usepackage{bm}
\usepackage{float}
\usepackage{hyperref}
\usepackage{color}
\usepackage{epstopdf}
\usepackage{cleveref}
\usepackage{braket}
\usepackage[svgnames]{xcolor}
\hypersetup{hidelinks,colorlinks=true,allcolors=DarkBlue}

\begin{document}

\preprint{APS/123-QED}

\title{Analysis of coherent quantum cryptography protocol \\ vulnerability to an active beam-splitting attack}

\author{D.A. Kronberg}
\affiliation{Steklov Mathematical Institute of Russian Academy of Sciences, Moscow 119991, Russia}
\affiliation{\mbox{Faculty of Computational Mathematics and Cybernetics, Moscow State University, Moscow 119899, Russia}}

\author{E.O. Kiktenko}
\affiliation{Theoretical Department, DEPHAN, Skolkovo, Moscow 143025, Russia}
\affiliation{Bauman Moscow State Technical University, Moscow 105005, Russia}

\author{A.K. Fedorov}
\affiliation{Russian Quantum Center, Skolkovo, Moscow 143025, Russia}
\affiliation{Acronis Ltd, Moscow 127566, Russia}
\affiliation{LPTMS, CNRS, Univ. Paris-Sud, Universit\'e Paris-Saclay, Orsay 91405, France}

\author{Y.V. Kurochkin}
\affiliation{Russian Quantum Center, Skolkovo, Moscow 143025, Russia}

\date{\today}
\begin{abstract}
We consider a new type of attack on a coherent quantum key distribution protocol [coherent one-way (COW) protocol]. 
The main idea of the attack consists in measuring individually the intercepted states and sending the rest of them unchanged. 
We have calculated the optimum values of the attack parameters for an arbitrary length of a channel length and compared this novel attack with a standard beam-splitting attack.
\end{abstract}
\maketitle

\section{Introduction}

Breakthrough in methods of manipulation for individual quantum systems plays a crucial role for the implementation of quantum technologies. 
Quantum technologies have a significant potential for design of computers~\cite{Feynman,Feynman2,Deutsch} and communications devices~\cite{BB84}.\,The 
use of quantum systems in the role of basic structural elements in computer allows one to achieve considerable improvement in a number of tasks \cite{Deutsch} 
such as search in unstructured databases~\cite{Grover}, and in integer factoring and discrete logarithm problems~\cite{Shor}. 
The last problems are of significant importance for public-key cryptography~\cite{DiffieHellman,Rivest}, 
which is based on the complexity of these tasks for classical computers.
But these tasks can be solved faster using quantum computers. 
This endangers the existing methods of information security using cryptographic tools. 
Furthermore, the absence of an efficient non-quantum algorithm for solving these problems still remains unproved.

In fact, appearance of quantum computing device limits possible tools for cryptography by two methods.
The first is to use problems without both classical and quantum efficient algorithms as novel public-key cryptography schemes.
These methods are in foundations  of post-quantum cryptography~\cite{Bernstein}. 
However, it is an open problem whether it is possible to design an effective algorithm that solves these problem, so the post-quantum cryptography keeps to be potentially vulnerable.

Another possible solution is to use cryptography with a private key. 
On the one hand, these systems (at certain conditions) are absolutely secure~\cite{Shannon}. 
If legitimate users (Alice and Bob) have identical random private keys, which are used only once, with the size being equal or greater than the size of the message, 
then according to the Shannon theorem messages encrypted using such key via the one-time pad scheme~\cite{Vernam} cannot be decrypted by an eavesdropper (Eve).
On the other hand, however, distribution of keys satisfying these demands is challenging.

\newpage

Towards the solution of this problem one can use the resource of quantum systems~\cite{BB84}.\,By transmitting information using individual quantum objects ({\it e.g.}, single photons) 
security is guaranteed by the principles of quantum physics~\cite{Gisin,Scarani}.
First is that the arbitrary quantum state cannot be cloned according to the no-cloning theorem~\cite{Wootters}. 
Second reason is that the pair of non-orthogonal states cannot be discriminated with unit probability.
Furthermore, any measurement leads to distortion or annihilation of a quantum state.
Thus, by distribution of a key with the use of single photons one can obtain a setup, which allows observing of inception in the key generation process.
Consequently, this opens a possibility to design a novel architecture for telecommunications systems, in which information security is guaranteed by the fundamentals laws of physics. 
Nevertheless, imperfections of technical realizations of key distribution systems such as absorption of photons in an optical fiber, 
efficiency of single-photon detectors, and possible actions of an eavesdropper exploiting these vulnerabilities lead to possible attacks.
In particular, if the length of a quantum channel exceeds a certain value then it is impossible to ensure confidentiality of key distribution~\cite{Scarani}. 

It should be noted that attacks on quantum key distribution systems can be divided into two classes. 
The first class is attacks on quantum key distribution protocols.
It is commonly understood that the quantum key distribution protocol is a general scheme of preparation and measurement of quantum states, 
and procedure of obtaining of a key on Alice and Bob sides from results of measurements of quantum states. 
The first quantum key distribution protocol is BB84.
When considering the attacks of this class, it is traditionally assumed~\cite{Gisin,Scarani} that an eavesdropper has all technological resources,
which do not contradict the physic laws,
such as quantum computer, quantum memory, and an ideal channel for transmitting of quantum states. 
Of course, the practical realization of the considered attack essentially depends on the required technological resources.
The second class presumes attacks on hardware realization of quantum-cryptographic systems (so-called ``quantum hacking'')~\cite{Makarov,Makarov2,Makarov3}.
For example, this is attacks on particular type of single-photon detectors~\cite{Makarov3}.

Quantum key distribution systems are available on the market. 
Nevertheless, the way to their implementation encounters with a number of technological challenges. 
One of the most promising methods of operating with quantum systems used in commercial devices consists of using coherent states of light for quantum key distributions 
(such as coherent one-way protocol, COW)~\cite{Stucki,Stucki2,Stucki3}.
These protocols are inspired by classical communications with optical fiber systems~\cite{Stucki3}.
The most significant advantage of the COW protocol is simplicity of its realization~\cite{Scarani,Stucki,Stucki2,Stucki3},
which is related to its simple optical scheme. 
This protocol is easy to implement in experiment, and it has been used in the project on the design of European network for quantum key distribution SECOQC~\cite{SECOQC}.
However, despite prevalence of this method and its high practical importance, 
an analysis of the possibility of attacks, which give to an eavesdropper an opportunity to obtain information about a quantum key, 
is an important task~\cite{Stucki,Stucki2,Stucki3,SECOQC,Branciard,Branciard2,Curty,Kronberg,Molotkov}.
We note that lower bounds on the key generation rate of a variant of the coherent-one-way quantum key distribution protocol have been obtained in Ref.~\cite{Moroder}.

We note that quantum key distribution systems are not communication systems in the full sense. 
The quantum resource in the form of single photons is used not for information transmitting, but for generation of a random identical for legitimate users sequence of bits (key).
The typical generation rate is such systems is about 10 kbit/s on the distances 50--80 km. 
It is a limitation of such systems that for generation of keys for legitimate users the direct channel is needed ({\it i.e.}, ``point-to-point'' network topology).
After distribution of a keys, it can be used for encryption on the one-time-pad regime or as a source of entropy. 
The final speed of information transmitting in such a case depends on the speed of the communication system transmitting encrypted information.

In this work, we consider a novel attack on the coherent quantum key distribution protocol. 
The key idea of the proposed attack is using of individual measurements of intercepted states and transfer of the remaining part in an unmodified form.
The suggested attack belongs to the first class (attacks on the protocol). 
One of its advantages is the fact that realization of this attack does not require using of the quantum memory or sophisticated elements,
but it is limited to a common assumption that an eavesdropper has a channel without losses. 
In the rest, the suggested attack has rather simple optical scheme for its realization,
allowing one to gain an advantage in comparison with the known beam-splitting attack under certain restrictions on the parameters of a key distribution system.  

\newpage

\section{COW protocol}

In the COW protocol for quantum key distribution Alice and Bob 
use a coherent state $|{\alpha}\rangle$ in a one of two time slots, whereas the vacuum state is in another window, to encode two information states with bit values $0$ and $1$.
Therefore, the corresponding states can be presented in the following form:
\begin{equation}\label{eq:inf_states}
\begin{split}
	|\psi_0\rangle=|\alpha\rangle\otimes|0\rangle, \quad
	|\psi_1\rangle=|0\rangle\otimes|\alpha\rangle,
\end{split}
\end{equation}
where $\mu = |\alpha|^2$ stands for the intensity of the coherent state $|{\alpha}\rangle$.

A straightforward scenario of attack in quantum cryptography is the ``intercept and resend'' attack. 
In this attack, Eve tries to measure the state in each time slot and resent of the resulting state (with amplification compensating losses).
In order to detect an eavesdropper, using this attack, in addition to the information states Alice and Bob employ the decoy states of the following form:
\begin{equation}
	|\psi_c\rangle=|\alpha\rangle\otimes|\alpha\rangle.
\end{equation}
The decoy states are used to detect attempts to distinguish between information states. 
During the attack an eavesdropper occasionally sends the information states instead the decoy states that would allow to detect it.
The fraction of the decoy states is denoted as $f$, it is typically about $10\%$.

\section{Beam-splitting attack \\ on the COW protocol}

In real optical fibers there is attenuation of signals leading to the fact that Bob obtains states of lower intensity.
The intensity of states obtained by Bob is as follows:
\begin{equation}\label{eq:Bob_intensity}
	\mu_B=\mu 10^{-\frac{\delta l}{10}},
\end{equation}
where $l$ is the length (in km), and $\delta$ is the attenuation coefficient (typical value of the attenuation coefficient for optical fibers is about $0.2$ dB/km).

In an analysis of attacks on the quantum key distribution protocol one can assume that Eve has unlimited technological resources. 
That is why one of the possible scenario is to use a beam splitter together with ideal identical channel (\emph{e.g.} by means of quantum teleportation).
Since states transform in a self-similar way both on a beamsplitter and in optical channel with attenuation, 
Eve can take a part of a state using beam-splitting with sending of the remaining part to Bob via channel replaced on the ideal one.

Maximal value of the intensity that Eve can take has the following form:
\begin{equation}\label{eq:Max_Eve_intensity}
	\mu_E^{\max}=\mu-\mu_B=\mu(1-10^{-\frac{\delta l}{10}}).
\end{equation}
From viewpoint of Bob, this modification of transferred states using a beamsplitter does not differs from losses, and it cannot be detected.
Further strategy of Eve is different for various attacks.

We consider the scenario of a beam-splitting attack, in which Eve stores the withdrawn states in order to then extract information~\cite{Kronberg}.
The set of states and measurements on the receiver side is such that if there is no Eve 
(as well as in the case of its withdrawal of a fraction of state and sending using a lossless channel) 
the Bob has zero quantum bit error rate (QBER), and the mutual information between Alice and Bob is equal to 1 after discarding inconclusive outcomes. 
Since Eva has less information, then in this attacks she introduces additional errors in order to make her information equal to the Bob information. 
Thus, information states of Eve in this attack are as follows:
\begin{equation}
	|\psi_0^E\rangle = |\sqrt{\mu_E^{\max}}\rangle\otimes|0\rangle, \quad
	|\psi_1^E\rangle = |0\rangle\otimes|\sqrt{\mu_E^{\max}}\rangle.
\end{equation}

The information, which can be extracted from such states by Eve at a collective measurement, is given by the Holevo quantity ($\chi$-value)~\cite{Holevo}:
\begin{equation}
	I_{AE}=\chi(\{|\psi_0^E\rangle, |\psi_1^E\rangle\}),
\end{equation}
It is easy to find a critical value of QBER, which Eve should add in order to make her information equal to Bob's information.  
The critical value of the QBER can be expressed via the binary entropy function as follows: 
\begin{equation}
	I_{AB}=1-h_2({\rm QBER}),
\end{equation}
where $h_2({\rm QBER})$ is the binary entropy function. 

The strong side of the beam-splitting attack is using of the collective measurements over the entire transmitted sequence, 
that allows one to achieve the $\chi$-value due the quantum superadditivity and obtain the theoretical maximum of information from these states.
However the disadvantage of such an attack is that the critical value of the QBER does not tend to zero at large lengths of the channel since the Eve information 
is limited by the $\chi$-value of initial states. 
At the same time, the longer is the channel, the higher are the losses, 
and one can use this fact by blocking some of the states, whose information had not been extracted.

\section{Active beam-splitting attack on the COW protocol}

\begin{figure*}
	\includegraphics[width=1\linewidth]{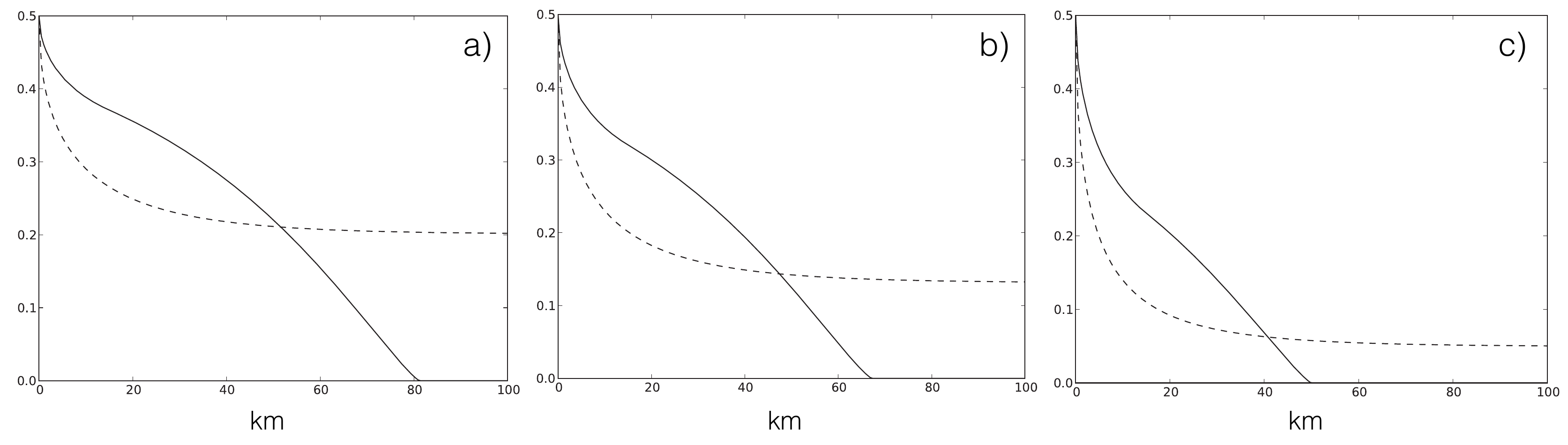}
	\vskip -3mm
	\caption
	{
	The critical value of the QBER for the standard beam-splitting attack (dashed line) and the considered active beam-splitting attack (solid line) for three different values of intensity 
	$\mu=0.1$ (a), 
	$\mu=0.2$ (b), 
	and $\mu=0.5$ (c) are shows as function of the channel length.
	}
	\label{fig:1}
\end{figure*}

In the present work, we suggest an alternative scenario to attack the COW protocol.
The difference from the beam-splitting attack is as follows.
First, in the considered attack Eve withdraws smaller part of states rather than in the beam-splitting attack, but still needs channel with lower losses. 
Second, the eavesdropper performs individual measurements. 
In the case of the inconclusive results of measurements ({\it i.e.}, at the detection of vacuum states in both positions) Eve is able to block state since the information is inaccessible. 
At the same time, in a general case Eve is unable to block all such states due to the fact that in this case the intensity on the receiver side is lower than it is expected, 
which discovers Eve.
Therefore, in order to estimate the efficiency of this attack we consider the idea similar to that in the beam-splitting attack, {\it i.e.}, 
in the cases where the length of the channel makes it impossible for Eve to block all states, of which information is not extracted, 
she introduces errors to make the Bob information equal to the Eve information. 
We refer this attack as the active beam-splitting attack since its using depends on ability of extracting information from withdrawn states by Eve. 

We describe details of actions of Eve with taking into account all the parameters. 
Eve withdraws a fraction of states with intensity $\mu_E$ and implements measurements of each time slot of the state.
The measurement in each time slot is described by the following observable: 
\begin{equation}
	M_0=|0\rangle\langle 0|,
	\quad 
	M_1=\sum_{i=1}^\infty |i\rangle\langle i|.
\end{equation}
The probability to obtain the result 1 at the measurement of the state with intensity $\mu_E$ reads
\begin{equation}\label{eq:Eve_conclusive_probability_inf}
	p_{\mathrm{conc}}^{\mathrm{inf}} = 1 - e^{-\mu_E}.
\end{equation}
This expression gives the probability to obtain the conclusive result at the measurement of the information state (\ref{eq:inf_states}) but with lower intensity. 
At the transmitting of the decoy state, the probability of the conclusive result is as follows:
\begin{equation}\label{eq:pcont}
	p_{\mathrm{conc}}^{\mathrm{cont}}=2e^{-\mu_E}(1 - e^{-\mu_E}) + (1 - e^{-\mu_E})^2.
\end{equation}
Thus, the total probability of the conclusive result by Eve taking into account the fraction of the decoy states is the following:
\begin{equation}
\label{Eve_conclusive_probability}
	p_{\mathrm{conc}}^E=(1-f)\times{p_{\mathrm{conc}}^{\mathrm{inf}}}+f\times{p^{\mathrm{cont}}_{\mathrm{conc}}}.
\end{equation}
It should be noted that the use of the decoy states does not allow to detect Eve since she still sends a fraction of states without measurements. 
Possible errors due to wrong distinctions of the information and the decoy states does not lead do decrease of the Eve information,
because the recordings about decoy states do not participate in the generation of the sifted key. 

In the case of inconclusive result of Eve measurement, the probability of which is equal to $1-p_{\mathrm{conc}}^E$, Eve seeks to block states transmitting to Bob.
In a general case, Eve is able to implement this not universally but for a small fraction of states to make this undetectable on the receiver side due to the large attenuation.
We calculate this fraction of states, which can be blocked. 

Bob expects states with intensity (\ref{eq:Bob_intensity}).
For these states the probability to obtain the conclusive result for information states is equal to $1-e^{-\mu_B}$.
In the reality, by using a beamsplitter Eve can keep for Bob states of higher intensity in order to block useless states, then the intensity of Bob is $\mu'_B=\mu-\mu_E$.
Consequently, permissible fraction of states $b$ blocked by Eve can be obtained from the relation
\begin{equation}\label{eq:block_probability}
	(1-b)(1-e^{-\mu_B'}) = 1 - e^{-\mu_B}.
\end{equation}
The fraction of blocked states can be determined for a given channel length and parameters of the Eve attack, in particular, the intensity of withdraw states. 
This intensity varies from zero to the maximal intensity $\mu_E^{\max}$ given by expression (\ref{eq:Max_Eve_intensity}). 
For each channel length Eve is able to chose the part of withdraw states maximizing her information. 

The Eve's information can be calculated as follows. 
The information is equal to unity for the cases, where Eve has conclusive results.
At the same time, for an arbitrary channel length Eve cannot block all states, where she extracted no information, then she still should send them to Bob. 
For the probability of the conclusive result and the probability of state block $b$, determined be Eq.~(\ref{eq:block_probability}), we have 
\begin{equation}\label{eq:Eve_information}
	I_{AE}=\frac{p_{\mathrm{conc}}^{\mathrm{inf}}}{1 - b}.
\end{equation}
One can see that if Eve is able to block all states with the inconclusive result of the measurement ({\it i.e.}, if $b=1-p_{\mathrm{conc}}^{\mathrm{inf}}$), then her information is equal to unity.
It is useless to consider the case, where the value $b$ is greater that the probability of the inconclusive result of the measurement, 
whereas at lower values Eve information decreases with the minimum at the case of absence of states block, which is equal to the probability of the conclusive measurement result, 
and it is the capacity of the channel with the attenuation. 

We consider a question about choice of the optimal value of the intensity of withdraw signal $\mu_E$ maximizing the withdraw information $I_{AE}$. 
By using expression (\ref{eq:Eve_conclusive_probability_inf}) and expression (\ref{eq:block_probability}) defining the value of the fraction of blocked by Eve states $b$, 
we then rewrite (\ref{eq:Eve_information}) as follows:
\begin{equation}\label{eq:Eve_information2}
	I_{AE}=\frac{\left(1-e^{-\mu+\mu_E}\right)\left(1-e^{-\mu_E}\right)}{1-e^{\mu_B}}.
\end{equation}
It is clear that $I_{AE}$ in a convex function of the argument $\mu_E$, which has a maximum at $\mu_E=\mu/2$.
However, since the intensity $\mu_E$ is bounded by the value $\mu_E^{\max}$ the maximum value of the withdraw information is achieved at
\begin{equation}
	\mu_E=\min\left(\mu_E^{\max},\mu/2\right)
\end{equation}
We note that in the case $\mu_E=\mu_E^{\max}$ we have $\mu_E^{\max}=\mu_E'$ and $b=1$. 
Thus, at $\mu_E<\mu/2$ the optimal strategy for Eve is to do not block states at all and send all messages to Bob. 

The critical length $l_\mathrm{crit}$at which it is reasonable for Eve to start block states is defined by the expression,
\begin{equation}
1-10^{-\frac{\delta{l_\mathrm{crit}}}{10}}=1/2,
\end{equation}
and it can be calculated as follows:
\begin{equation}
	l_\mathrm{crit}=10\log_{10}2/\delta=3/\delta.
\end{equation}
For a typical value $\delta=0.2$ dB/km, we then have $l_\mathrm{crit}=15$ km.

\section{Discussions of results}

In conclusion, we one more time briefly describe a scenario of the considered attack and the method of finding of the critical QBER in the protocol.
For a given value of the channel length, Eve calculates the maximal intensity of states, which can be withdrawn. 
Then, she considers possibility of withdraw of states with different intensity from zero to the maximal value.
For each state one can calculate the fraction of states, which Eve can block, and the Eve information. 
The greater is the intensity of the state received by Bob, the larger fraction can be blocked.
Eve chooses the intensity maximizing her information. 
If she can block all states with inconclusive results of measurements, Eve is able to attack without introducing of additional error, and the protocol is totally insecure. 
However, if the fraction of states, which can be blocked, is lower, 
then Eve incorporates errors in the channel between Alice and Bob in order to offset the difference between Bob's and her information.
The minimum value of errors, whereby their information are equal to each other, is a critical value of the QBER of the protocol against this attack.
The critical value of QBER for the standard beam-splitting attack and the considered active beam-splitting attack for three different values of intensity ($\mu=$ 0.1, 0.2, and 0.5)
are shows as function of the channel length in Fig.~\ref{fig:1}.

\begin{figure}
	\begin{center}
	\includegraphics[width=0.8\linewidth]{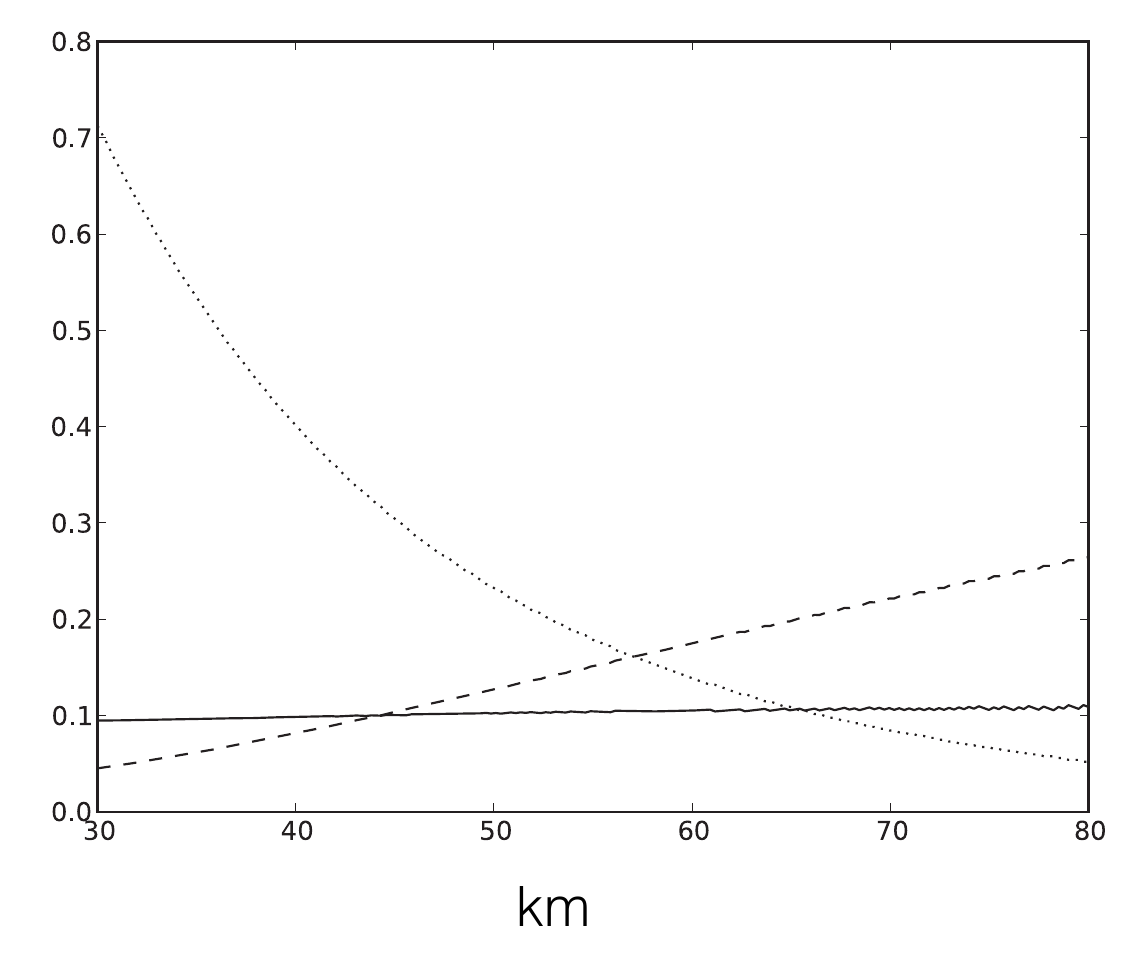}
	\end{center}
	\vskip -7mm
	\caption{The optimal intensity in a presence of active beam-splitting attack for different channel lengths (dotted line), 
	the critical QBER at the optimal choice of the intensity by Eve (solid line), and the critical QBER of the corresponding beam-splitting attack at the same intensity (dashed line).}
	\label{fig:2}
\end{figure}

The question may arise as to which the intensity of the initial state is optimal for legitimate users under the assumption that the eavesdropper uses this particular attack. 
On the one hand, it is clear that in order to increase the critical QBER they have to take the lowest intensity of the initial states. 
On the other hand, low intensity of the initial states leads to low key generation rate because of too large proportion of too many states are lost.

We can find the optimal intensity as follows.
Assume that Eve uses withdraw of the maximum fraction of states, however without incorporation of errors in the communication channel between Alice and Bob 
(since the value of errors is not the mark). 
The length of secret key of Alice and Bob recalculated on a single state 
is given by difference between the information between Alice and Bob $I_{AB}$ and the information between Alice and Eve $I_{AE}$.
The first is equal to the capacity of the erasure channel with erasure probability equal to  $e^{-\mu_B}$, the second can be calculated by using Eq.~(\ref{eq:Eve_information})
and then multiplying on the probability of the conclusive result of Bob. 
Therefore, the optimal value of the intensity for the given value of the channel is those, which maximizes the difference
\begin{equation}\label{optimal_intencity}
	I_{AB}-I_{AE}=1-e^{-\mu_B}-(1-e^{-\mu_B})\frac{p_{\mathrm{conc}}^{\mathrm{inf}}}{1-b}.
\end{equation}

Fig.~\ref{fig:2} shows the optimal intensity for different channel lengths and the critical value at the optimal choice of the intensity if Eve uses this attack. 
For a comparison, we add the critical value of QBER for the beam-splitting attack at the fixed length and given choice of intensity, 
however, we note that the initial intensity is optimized on the basis of the assumption that the eavesdropper uses the active beam-splitting attack.

The disadvantage of this attack is that Eve is unable to amplify the signal transmitting to Bob for the states, where she has the conclusive results.
Another weak side of the attack is that Eve has to measure at once, which it eliminates the possibility of achieving of the superadditive information. 
On the other hand, in contrast to the usual attacks with a beam splitter, such the attack, starting with the critical channel length, is possible without introducing of additional errors. 
However, the attack with unambiguous measurements~\cite{Molotkov} is more effective since it is possible to enhance the intensity of the premises from which all the information can be extracted. 
The advantages of the considered attack can also be attributed relatively simple technical implementation.

Thus, the discussed active beam-splitting attack is interesting primarily as implemented in the current technological level than the optimum in terms of the eavesdropper, 
limited only by the laws of physics. 
The development of the idea of unchanged forwarding states seems to be topical in the context of other quantum-cryptography protocols that are based on the use of coherent states.

We note also that despite the fact that the considered attack does not lead to a change of the status type (decoy state can not be transformed into a signal, and vice versa), 
it leads to a distortion of the statistics getting decoy or information states. 
This phenomenon is due to both the change in the intensity of the pulses sent to Bob (by $\mu_B$ or $\mu_B'$), 
and that the blocking probability able to determine the probability of getting conclusive result by Eve.
In turn, in the case of this decoy state likelihood is higher than in the case of the signal, and the total probability of obtaining the decoy state increases. 
As a result, the attack can be considered potentially registered by taking into account statistical registration of decoy states in the classical post-processing key. 
Usually, however, the protocol is considered a requirement of the absence of conditions such as change, 
and such registration statistics require a significant change in the protocol in terms of evaluation of the interceptor information.

An interesting and relevant question is about required modification of the COW protocol for taking into account the proposed attack.
However, this question further analysis, and it is beyond the scope of this work.

\section{Conclusion}

In this paper, we considered a new type of attack on a coherent quantum key distribution protocol COW with the use of an active beam splitter. 
The optimum values of the parameters of attack for an arbitrary length of the channel were calculated, and the comparison with a standard beam-splitting attack was performed .

The advantage of the considered attack is rather simple technical implementation.
It should be noted that the suggested attack is actual, in fact, for channels of arbitrary length.
However, it is of special interest for quantum-cryptographic systems operating in an urban environment and using short (30--50 km) urban fiber-optic communication lines with fairly high losses.
In recent experiments on quantum key distribution in urban conditions, 
losses in the channel of 30 km length have been on the level of 11 dB at the key generation rate on the level of 0.5 kbit/s~\cite{Miller} after post processing~\cite{Kiktenko}.

\newpage

We are grateful to A.S. Trushechkin and O.V. Lychkovskiy for helpful discussions, and to the reviewers for valuable comments.

The financial support from Ministry of Education and Science of the Russian Federation in the framework of the Federal Program 
(Agreement 14.579.21.0105, ID RFMEFI57915X0105) is acknowledged.

\end{document}